\documentclass{birkjour}
\usepackage{amsmath}
\usepackage[T1]{fontenc}
\usepackage[latin9]{inputenc}
\usepackage{verbatim}
\usepackage{amsbsy}
\usepackage{amstext}
\usepackage{amsthm}
\usepackage{amssymb}
\usepackage{esint}
\usepackage{ulem}
\usepackage{mathtools}
\usepackage{ulem}
\usepackage{graphicx}
\usepackage[usenames,dvipsnames]{color}
\usepackage{longtable}
\usepackage{dcolumn}
\usepackage{nicefrac}
\usepackage{babel}
\usepackage{amsfonts}
\usepackage{babel}

\PassOptionsToPackage{normalem}{ulem}
\makeatletter
\numberwithin{equation}{section}
\numberwithin{figure}{section}
\theoremstyle{plain}

\theoremstyle{definition}

\theoremstyle{remark}

\theoremstyle{plain}

\providecommand{\U}[1]{\protect\rule{.1in}{.1in}}
\numberwithin{equation}{section}
\numberwithin{figure}{section}
\theoremstyle{plain}
\makeatother
\providecommand{\corollaryname}{Corollary}
\providecommand{\definitionname}{Definition}
\providecommand{\remarkname}{Remark}
\providecommand{\theoremname}{Theorem}

\begin{document}

\title{Quantization of pseudo-hermitian systems}
\author{M.C. Baldiotti \and R. Fresneda}
\email{balddioti@uel.br}
\email{rodrigo.fresneda@ufabc.edu.br}

\begin{abstract}
This work is a generalization of \cite{baldiotti2021} to Grassmann algebras
of arbitrary dimensions. Here we present a covariant quantization scheme for
pseudoclassical theories focused on non-hermitian quantum mechanics. The
quantization maps canonically related pseudoclassical theories to
equivalent quantum realizations in arbitrary dimensions. We apply the
formalism to the problem of two coupled spins with Heisenberg interaction.
\end{abstract}

\maketitle

\address{Londrina State University - UEL - Brasil}

\address{Federal University of ABC - UFABC - Brasil}

\section{Introduction}

Among the postulates of quantum mechanics, there is one which demands that
operators related to measurable physical quantities be self-adjoint. This is
to ensure that the spectrum of such operators are real, and to relate these
real values to measurable quantities. However, self-adjointness is a
sufficient, albeit not necessary condition for an operator to have a real
spectrum. Notable examples were given in \cite{Bender98, Fern}.
This fact, which seems like a mathematical peculiarity, has practical
applications, especially in optical physics \cite{Regensburger, Bittner,
Chong, Peng}, and its origin is related to PT (Parity-Time) symmetry \cite{Batal, Ali2003}.

Another feature we would like to point out is the freedom in realizing the
algebra of operators during the quantization process. At first, this freedom
might lead to an ambiguity of the physical description, since one might end
up with theories describing different physical processes. This problem was
elucidated by Stone and von Neumann \cite{rs}, who show that these quantum
theories which arise according to different realizations of the operator
algebra are linked by unitary operators. Thus, they are equivalent to the
extent of the measurable behavior of the system they describe.

These two points stress the importance of the notions of self-adjointness
and unitarity. Restricting our discussion to systems with finite degrees of
freedom\footnote{%
For systems with finite degrees of freedom, the notion of self-adjointness
of a linear operator coincides with the notion of hermiticity.}, such
concepts are directly related, because unitary operators are obtained
through exponentiation of hermitian matrices (up to the imaginary unit).
Furthermore, in the context of complex matrices, a hermitian matrix is
usually defined as one which is equal to its transposed conjugate. However,
the existence of non-hermitian operators with real spectrum (which can thus
represent physical quantities) raises doubt on the necessity of the
postulate we described. On the other hand, the \textquotedblleft
non-orthogonality\textquotedblright\ of the eigenvectors of such operators
seems to go against a spectral decomposition allowing a probabilistic
interpretation. However, the notion of orthogonality derives from the
definition of the inner product in the Hilbert space which realizes the
quantum description. Thus, it is clear that the problem of non-orthogonality
can be solved by the introduction of a new inner product \cite%
{Ali2002-1,Ali2002-2,Ali2002-3,Ali10}. As a result, instead of doing away
with the hermiticity (or self-adjointness) postulate, we only need to
redefine hermiticity.

There still remains the question about the unitarity of operators. For
non-hermitian operators, that is, operators which are not hermitian
according to the usual inner product (or canonical inner product),
exponentiation leads to non-unitary operators, so it was believed that the
unitarity of the evolution operator is lost \cite{Yac04}. We show, however,
in \cite{baldiotti2021}, that this is not true, as long as one adopts a
general notion of unitarity or isometry. More specifically, this notion is
related to the isometry between Hilbert spaces with different inner
products. In \cite{baldiotti2021}, we show that, for systems with two
degrees of freedom, the appearance of different inner products is associated
to the possibility of relating the corresponding pseudoclassical
descriptions by canonical transformations. As a physical application, we
considered the case of spin precession under a magnetic field for a
two-level system, that is, the quantum mechanics of a spin-$1/2$ particle
interacting with an external magnetic field, disregarding the spatial
dynamics. Among the exact solutions of the latter, we highlighted the Rabi
problem \cite{Rabi2,Rabi},

By adding an imaginary contribution to the magnetic field, we constructed a
damped version of the Rabi problem. One can fine-tune this imaginary
contribution such that the Rabi problem is still described by a hermitian
Hamiltonian. This was the main physical result in \cite{baldiotti2021}, and
it lead to some practical applications.

The goal of this work is to extend the results obtained in \cite%
{baldiotti2021} to systems with arbitrary finite number of degrees of
freedom. Besides the theoretical relevance, such a generalization allows the
exploration of new physical properties, such as the interaction of two or
more spin-$1/2$ particles.

On a formal level, the main idea can be summarized in the following diagram.
Given the space $\mathcal{L\left( H\right)}$ of  operators on a finite-dimensional Hilbert space $%
\mathcal{H}$, one constructs a quantization map $Q:M\rightarrow \mathcal{\
L\left( H\right) }$ defined on a $\mathbb{Z}_{2}$-graded symplectic space $M$
such that the diagram commutes:

\begin{figure}[h]
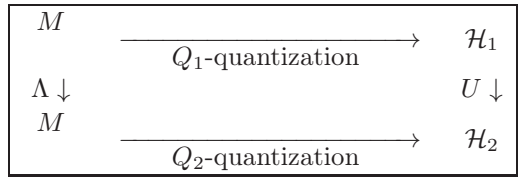

\begin{equation*}
\begin{array}{|ccc|}
\hline
\begin{array}{c}
\text{$M$} \\ 
\\ 
\end{array}
& 
\begin{array}{c}
\\ 
\overrightarrow{\text{\ \ \ \ \ \ }Q_{1}\text{-quantization \ \ \ \ \ \ }}%
\end{array}
& \mathcal{H}_{1} \\ 
\Lambda \downarrow &  & U\downarrow \\ 
\begin{array}{c}
\text{$M$} \\ 
\\ 
\end{array}
& 
\begin{array}{c}
\\ 
\overrightarrow{\text{\ \ \ \ \ \ $Q_{2}$-quantization \ \ \ \ \ \ }}%
\end{array}
& \mathcal{H}_{2} \\ \hline
\end{array}%
\end{equation*}%
\caption{covariance diagram}
\label{fig:diagram}
\end{figure}

The map $\Lambda$ is a linear canonical transformation (linear
symplectomorphism) on $M$, and $U:\mathcal{H}_{1}\rightarrow\mathcal{H}_{2}$
is a linear isomorphism between Hilbert spaces with inner products $%
\left\langle \cdot,\cdot\right\rangle _{1}$ and $\left\langle
\cdot,\cdot\right\rangle _{2}$, respectively (so $\left\langle
x,y\right\rangle _{1}=\left\langle Ux,Uy\right\rangle _{2}$ $\forall x,y\in%
\mathcal{H}_{1}$).

Now the ideology behind this construction is the following. One is sometimes
faced with non-hermitian observables in Quantum Mechanics (see \cite{Ali10}%
), that is, nonsymmetric operators which are related to some physical
property. Let's say operator $O$ on $\mathcal{H}_{1}$ is not symmetric on $%
\mathcal{H}_{1}$. Then the question naturally arises as to whether this
operator is symmetric on another Hilbert space $\mathcal{H}_{2}$ of the same
dimension. If this is the case, then we say that $O$ is pseudo-hermitian (or
pseudo-symmetric)\footnote{%
These operators are also called crypto-hermitian or quasi-hermitian.}.

In the event that $O$ is pseudo-hermitian, one is able to obtain a linear
isomorphism $U $ mapping the Hilbert spaces $\mathcal{H}_{1}$ and $\mathcal{H%
}_{2}$ and thus obtain an equivalent physical description of the same
physical system. Even though $O$ is not symmetric on $\mathcal{H}_{1}$, and
it is thus not considered a physical observable in usual Quantum Mechanics,
the operator $U^{-1}OU$ is symmetric on $\mathcal{H}_{1}$ and it therefore
suits the standard quantum mechanical framework.

By means of this construction, we are able to establish to what extent one
is able to fine-tune the physical system in order to maintain
pseudo-hermiticity. In other words, the covariant quantization scheme we
propose can tell us what are the parameter domains for which one can still
go over to a standard quantum mechanical description.

In section \ref{quant} we consider the quantization map between the
pseudoclassical phase space and the Hilbert space of Quantum Mechanics.
Finally, in section \ref{realization} we illustrate the formalism with a
physical setup of two coupled spins in the Heisenberg interaction.

\section{Covariant quantization and hermiticity}

\label{quant}

First, we will simply state some definitions and results in pseudoclassical
mechanics for nonrelativistic systems \cite{Cas76,Cas762,Ber77,gitman1991},
following our work \cite{baldiotti2021}. Generally speaking,
pseudoclassical models are $\mathbb{Z}_{2}$-graded symplectic spaces which
function as classical analogs for systems with fermionic degrees of freedom.

Given a Grassmann algebra $G_n(\xi)$ whose generators satisfy $%
\xi_{i}\xi_{j}+\xi_{j}\xi_{i}=0$ for $i=1,..,n$, one can define functions\footnote{%
Here and throughout sums over repeated indices are implied.} 
\begin{equation}
f(\xi)=f_{0}+f_{i}\xi_{i}+f_{ij}\xi_{i}\xi_{j}+\frac{\mathrm{i}}{3!}%
f_{ijk}\xi_{i}\xi_{j}\xi_{k}+...+\frac{1}{n!}f_{i_{1}\cdots
i_{n}}\xi_{i_{1}}\cdots\xi_{i_{n}}\,,  \label{general-f}
\end{equation}
where the coefficients are completely antisymmetric complex constants.

A pseudoclassical model is defined by specifying an action functional $%
S=\int L( \xi ,\dot{\xi})dt$, where the Lagrange function $L$ is a function
on $G_{n}(\xi )$. Our convention for the derivative operator $\partial
/\partial \xi _{i}$ is that it acts from the right. One can define a
symplectic structure and a phase space $M$ by introducing canonical momenta
in the usual way and by taking the natural $\mathbb{Z}_{2}$-graded bracket
in the coordinates $(\xi ,\pi )$ as 
\begin{equation}
\left\{ f,g\right\} =\frac{\partial f}{\partial \xi _{i}}\frac{\partial g}{%
\partial \pi _{i}}-\left( -1\right) ^{P_{f}P_{g}}\frac{\partial g}{\partial
\xi _{i}}\frac{\partial f}{\partial \pi _{i}}\,,
\end{equation}%
where $f$ and $g$ are functions of definite parity $P_f$ and $P_g$,
respectively. With the above bracket, one can obtain the Hamiltonian
function $H$.

For a general function $f(\xi )$ as in Eq.~(\ref{general-f}), we define an
involution $\ast :G_{n}(\xi )\rightarrow G_{n}(\xi )$ such that its action
on the generators $\xi _{i}$ is given by 
\begin{equation}
\xi _{i}^{\ast }=\xi _{i},\,i=1,2,...,n\,.  \label{star-involution}
\end{equation}%
Therefore, real elements in the algebra (those for which $f^{\ast }=f$) are
given by (\ref{general-f}) with 
\begin{equation}
f_{i_{1}\cdots i_{k}}=\left( -1\right) ^{\frac{k\left( k-1\right) }{2}}\bar{f%
}_{i_{1}\cdots i_{k}}~,
\end{equation}
where the bar denotes complex conjugation.

Suppose we consider a linear canonical transformation on the
pseudo-mechanical phase space, defined as a map $(\xi,\pi)\mapsto(\zeta,%
\varpi)$, which preserves the symplectic structure in that the only
nonvanishing brackets between the new coordinates are $\{\zeta_{i}(\xi,\pi),%
\varpi_{j}(\xi,\pi)\}=\delta_{ij}$.

In nonrelativistic spinning models (see \cite{baldiotti2021} and references
therein), due to the presence of constraints, we observe that $\pi$ is
proportional to $\xi$, so we write the linear canonical transformation
simply as 
\begin{equation}
\zeta_{i}=\Lambda_{ik}\xi_{k}~\text{and}~\varpi_{j}=\Lambda_{jl}\pi_{l}\,.
\label{ctc}
\end{equation}
Then, demanding that this transformation is canonical implies $\Lambda
\Lambda^{T}=I_n$, where $I_n$ is the $n\times n$ identity matrix. That is, $%
\Lambda$ is an orthogonal matrix. In principle, $\Lambda$ can have complex
entries, so $\Lambda\in O(n,\mathbb{C})$.

For a general monomial in (\ref{general-f}), the coefficients $%
f_{i_{1}\cdots i_{k}}$ behave as tensors under $O\left(n,\mathbb{C}\right)$, 
\begin{equation}
f_{i_{1}\cdots i_{k}}\mapsto\tilde{f}_{i_{1}\cdots
i_{k}}=\Lambda_{i_{1}j_{1}}\cdots \Lambda_{i_{k}j_{k}}f_{j_{1}\cdots
j_{k}}\,.  \label{eq:f-transf}
\end{equation}
Let us consider an involution $+:G_{n}\left(\zeta\right)$ in $%
G_{n}\left(\zeta\right)$ such that real elements in $G_{n}\left(\xi\right)$
are mapped by the linear canonical transformation (\ref{ctc}) to real
elements in $G_{n}\left(\zeta\right)$. The action of the involution is given
by 
\begin{equation}
\zeta_{i}^{+}=\zeta_{i}^{\ast}=\rho_{ki}\zeta_{k}\,,  \label{plus-involution}
\end{equation}
where $\rho=\Lambda \Lambda^{\dagger}$ and $\Lambda^{\dagger}$ is the
conjugate transpose of $\Lambda$. The $\zeta$-terms above are taken as
function of $\xi$. In fact, the general element of $G_{n}\left(\zeta\right)$
is $+$-real iff its complex coefficient satisfy 
\begin{equation}
f_{i_{1}\cdots i_{k}}=\left(-1\right)^{\frac{k\left(k-1\right)}{2}%
}\rho_{i_{1}j_{1}}\cdots\rho_{i_{k}j_{k}}\bar{f}_{j_{1}\cdots j_{k}}\,.
\end{equation}
Moreover, due to (\ref{eq:f-transf}), it follows that the real elements of $%
G_{n}\left(\xi\right)$ are mapped to real elements of $G_{n}\left(\zeta%
\right)$ by the orthogonal transformation $\Lambda$, that is, $%
f=f^{\ast}\Leftrightarrow g=g^{+}$ where $f(\xi)=g(\zeta(\xi))$.

Now let us discuss a general quantization scheme for pseudoclassical
models. From here on we restore $\hbar $. Let us define a quantization map $%
Q:M\rightarrow \mathcal{L}(\mathcal{H}_{\eta })$, where $M$ is the pseudoclassical
symplectic with its underlying bracket and $\mathcal{L}(\mathcal{H}_{\eta })$ is the
set of linear operators on the Hilbert space $\mathcal{H}_{\eta }=(%
\mathbb{C}^{m},\langle \cdot ,\cdot \rangle _{\eta })$. The inner product 
\begin{equation}
\left\langle x,y\right\rangle _{\eta }:=\left\langle x,\eta y\right\rangle
\,,\,\,\forall x,y\in \mathcal{H}_\eta\,,  \label{eq:eta-product}
\end{equation}%
is indexed by the positive linear automorphism $\eta$, and $%
\left\langle\cdot,\cdot\right\rangle$ is the canonical inner product on $%
\mathbb{C}^m$.

It suffices to define the map on monomials, following the
anti-symmetrization rule 
\begin{equation}
Q\left(\xi_{_{1}}\xi_{_{2}}\cdots\xi_{_{n}}\right)=\frac{1}{n!}\sum_{\text{%
perm}}\left(-1\right)^{\sigma\left(\text{perm}\right)}Q\left(\xi_{i_{1}}%
\right)Q\left(\xi_{i_{2}}\right)\cdots Q\left(\xi_{i_{n}}\right)\,,
\label{symmetrization}
\end{equation}
and extend it linearly to all functions. Furthermore, the quantization map $%
Q $ is required to map the unit to the identity in $\mathcal{H_{\eta}}$, $%
Q(1)=\mathbb{I}$. In our case, the above requirements imply the following
for general classical functions: 
\begin{equation}
Q\left(f\right)=f_{0}\mathbb{I}+f_{i}Q(\xi_{i})+f_{ij}Q(\xi_{i})Q(\xi_{j})+%
\cdots+\frac{1}{n!}f_{i_{1}\cdots i_{k}}Q(\xi_{i_{1}})\cdots
Q(\xi_{i_{k}})\,.
\end{equation}

By requiring that the quantization $Q$ map the generators $\xi_{i}$ to
symmetric operators $Q\left(\xi_{i}\right)$, it follows that for a general
real element in $G_{n}\left(\xi\right)$ one has 
\begin{equation}
f=f^{\ast}\Rightarrow\langle x,Q\left(f\right)y\rangle_{\eta}=\langle
Q\left(f\right)x,y\rangle_{\eta}\,.
\end{equation}
That is, for real functions $f$, $Q(f)$ is symmetric, $Q^{*}(f)=Q(f)$. The
map $Q$ is also required to satisfy the correspondence principle 
\begin{equation}
\{f,h\}_{D(\phi)}=\lim_{\hbar\rightarrow0}\frac{1}{\mathrm{i}\hbar}\left[%
Q(f),Q(h)\right]]\,,
\end{equation}
where $\{\cdot,\cdot\}_{D(\phi)}$ is the Dirac bracket constructed from the
set $\{\phi\}$ of second-class constraints (see \cite{gitman1991}) and $%
[\cdot\,,\cdot]$ is a $\mathbb{Z}_{2}$-graded commutator: 
\begin{equation}
\left[Q(f),Q(h)\right]=Q(f)Q(h)-(-1)^{P_{f}P_{h}}Q(h)Q(f)\,,
\end{equation}
for all homogeneous functions $f$ and $h$. Thus, one has for the dynamical
variables $\{\xi_{i}\}_{i=1}^{n}$ the basic anti-commutation relations 
\begin{equation}
\left[Q(\xi_{i}),Q(\xi_{j})\right]=\hbar\delta_{ij}  \label{AC}
\end{equation}
for the Clifford algebra $Cl_{n}\left(\mathbb{C}\right)$. We shall thus
consider $\mathcal{H}_{\eta}$ to be a $2^{\left[n/2\right]}$-dimensional
unitary irreducible representation of $Cl_{n}\left(\mathbb{C}\right)$, i.e., 
$\dim\mathcal{H}_{\eta}=2^{\left[n/2\right]}$.

As in the classical case (\ref{ctc}), let us consider the canonical
transformation $\zeta_{i}=\Lambda_{ij}\xi_{j}$ with $\Lambda\in O(n,\mathbb{C%
})$. The quantization map $Q^{\prime}:M\rightarrow \mathcal{L}(\mathcal{H}_{\rho})$, $%
\mathcal{H}_{\rho}=\left(\mathbb{C}^{m},\rho\right)$, is now defined on a
function $g\left(\zeta\right)$ of the generators $\zeta_{i}$ of $%
G_{n}\left(\zeta\right)$ in an analogous manner. A natural question is then
what is the relation between $Q(f)$ and $Q^{\prime}(g)$, where $%
g(\zeta)=f(\xi(\zeta))$. Let $U:\mathcal{H}_{\eta}\rightarrow\mathcal{H}%
_{\rho}$ be a linear isomorphism. Then, the quantization procedure is
covariant if the diagram \ref{fig:diagram} commutes. This requires that the
images of the classical functions $g\left(\zeta\right)$ and $%
f\left(\xi\right)$ are related by the similarity transformation 
\begin{equation}
Q^{\prime}(g)=UQ(f)U^{-1}\,.  \label{similarity}
\end{equation}
The image of the generators $\zeta_{i}$ of $G_{n}\left(\zeta\right)$ by the
quantization map $Q^{\prime}$ satisfy the Clifford algebra relations $\left[%
Q^{\prime}\left(\zeta_{i}\right),Q^{\prime}\left(\zeta_{j}\right)\right]%
=\hbar\delta_{ij}$. This follows from considering the functions $%
f_i\left(\xi\right)=\Lambda_{ij}\xi_{j}$ and $g_i\left(\zeta\right)=%
\Lambda_{ij}\xi_{j}\left(\zeta\right)=\zeta_{i}$, so that $%
Q^{\prime}\left(\zeta_{i}\right)=\Lambda_{ij}Q\left(\xi_{j}\right)$, and
taking into account that $\Lambda\in O\left(n,\mathbb{C}\right)$. The
similarity relation (\ref{similarity}) preserves the canonical relation (\ref%
{AC}), and can be regarded as a quantum canonical transformation induced by
the classical canonical transformation (\ref{ctc}).

Note that given the linear isomorphism $U:\mathcal{H}_{\eta }\rightarrow 
\mathcal{H}_{\rho }$, the metric $\rho $ is given in terms of the metric $%
\eta $ and $U$ as $\rho =(U^{\dagger})^{-1}\eta U^{-1}$, where $U^{\dagger}$
is the usual adjoint, the conjugate transpose of $U$. In the applications
that will follow, we take $\eta $ to be the identity, so $\rho =(UU^{\dagger
})^{-1}$.

Moreover, let $Q^{\prime+}(g)$ denote the adjoint of $Q^{\prime}(g)$ with
respect to the inner product $\langle\cdot\,,\cdot\rangle_{\rho}$. Using (%
\ref{similarity}), it follows that $Q^{\prime+}(g)=UQ^{*}(f)U^{-1}$. Thus,
for $+$-real $g$, the corresponding operator is $\rho$-symmetric, $%
Q^{\prime+}(g)=Q^{\prime}(g)$, since real $g$ ($g^{+}=g$) implies real $f$ ($%
f=f^{\ast}$), and $Q^{*}(f)=Q(f)$. 

If $\tilde{U}:\mathcal{H}_{\eta}\rightarrow\mathcal{H}_{\rho}$ is another
linear isomorphism, then $\tilde{U}=SU$ for some $S$ which preserves $\rho$,
i.e., $\rho=S^{\dagger}\rho S$. Let $\tilde{Q}:M\rightarrow \mathcal{L}\left(\mathcal{H%
}_{\rho}\right)$ be another quantization map such that one has covariance
with isomorphism $\tilde{U}$, namely, $\tilde{Q}\left(g\right)=\tilde{U}%
Q\left(f\right)\tilde{U}^{-1}$. Then, the relation between $Q^{\prime}$ and $%
\tilde{Q}$ is $\tilde{Q}=SQS^{-1}$, that is, they are related by a $\rho$%
-preserving linear map. Thus, covariance holds up to $\rho$-preserving
transformations.

In particular, the quantization of the Hamiltonian function $%
H\left(\xi\right)$, seen as a real even element in $G_{n}\left(\xi\right)$,
will give the symmetric operator $\hat{H}_{\xi} \coloneqq Q\left(H\right)$.
On the other hand, the quantization of $H_{\zeta}\left(\zeta\right)=H\left(%
\xi\left(\zeta\right)\right)$ is such that, as in (\ref{similarity}), 
\begin{equation}
\hat{H}_{\zeta}\coloneqq Q\left(H_{\zeta}\right)=U\hat{H}_{\xi}U^{-1}\,,\,\,%
\hat{H}_{\zeta}=\hat{H}_{\zeta}^{+}\,,
\end{equation}
that is, it is $\rho$-symmetric, since $H_{\zeta}\left(\zeta\right)$ is $+$%
-real in $G_{n}\left(\zeta\right)$.

\subsection{The $n=3$ case}

For completeness, let's review the main aspects of the $n=3$ case. This case
is treated in detail, and with applications, in \cite{baldiotti2021}. The
pseudoclassical Hamiltonian can be obtained by requiring that $\xi_{i}$
transform as a vector under $O\left(3\right)$, 
\begin{equation}
H_{B}=-\frac{\mathrm{i}}{2}\varepsilon_{ijk}\xi_{i}\xi_{j}B_{k}\,,
\label{HSC}
\end{equation}
where $B_{k}$ transforms as a pseudo-vector (for instance, like the magnetic
field). Thus, the equation of motion for $\xi_{i}$ becomes 
\begin{equation}
\dot{\xi}_{i}=\{\xi_{i},H\}_{D(\phi)}=-\varepsilon_{ijk}\xi_{j}B_{k}\,,
\label{CEM}
\end{equation}
which is recognized as the classical precession equation, like a magnetic
moment immersed in a magnetic field $\mathbf{B}=\left(B_{1},B_{2},B_{3}%
\right)$. In particular, $H_{B}(\xi)$ is $\ast$-real when $\mathbf{B}\in%
\mathbb{R}^{3}$.

Furthermore, under transformation (\ref{ctc}) the Hamiltonian function (\ref%
{HSC}) becomes 
\begin{equation}
H_{F}\left(\zeta\right)=-\frac{\mathrm{i}}{2}\varepsilon_{ijk}\zeta_{i}%
\zeta_{j}F_{k}\,,  \label{zeta-hamiltonian}
\end{equation}
where 
\begin{equation}
F_{k}=(\det \Lambda)\Lambda_{kl}B_{l}\,.  \label{FfromB}
\end{equation}
Relation (\ref{FfromB}) implies that 
\begin{equation}
F^{2}=F_{i}F_{i}=\left(\det
\Lambda\right)^{2}\Lambda_{ij}\Lambda_{ik}B_{j}B_{k}=%
\delta_{jk}B_{j}B_{k}=B^{2}.  \label{Cond0}
\end{equation}
Thus, if $\mathbf{B}$ is a real field, then from the previous relation it
follows that $F^{2}$ is a positive real number for an arbitrary complex
field $\mathbf{F}$. Indeed, considering such a complex field $\mathbf{F}$,
it follows that $H_{F}\left(\zeta\right)$ is $+$-real.

The quantization $Q$ of a general element in $G_{3}\left(\xi\right)$ has the
following form 
\begin{equation}
Q(f)=f_{0}\mathbb{I}+f_{i}Q(\xi_{i})+f_{ij}Q(\xi_{i})Q(\xi_{j})+\mathrm{i}%
\frac{1}{3!}k_{f}\varepsilon_{ijk}Q(\xi_{i})Q(\xi_{j})Q(\xi_{k})\,,
\end{equation}
and we take $\eta$ to be the canonical inner product in $\mathbb{C}^{2}$. A
unitary representation of the Clifford algebra (\ref{AC}) on $\mathbb{C}^{2}$
is given by the Pauli matrices $\sigma_{i}$ as 
\begin{equation}
Q(\xi_{i})=\sqrt{\frac{\hbar}{2}}\sigma_{i}\,.
\end{equation}
Then, following the quantization rule (\ref{symmetrization}), the
Hamiltonian operator $\hat{H}_{B}\coloneqq Q(H_{B})$ (image of (\ref{HSC})
by the quantization map $Q$) is 
\begin{equation}
\hat{H}_{B}=\frac{\hbar}{2}\boldsymbol{\sigma}\cdot\boldsymbol{B}\,.
\label{HB}
\end{equation}
Given a realization of the algebra (\ref{AC}), it is immediate to write a
realization for the operators $Q(g(\zeta))$ using relation (\ref{similarity}%
). For the particular case of the Hamiltonian function $H_{F}$ in Eq. (\ref%
{zeta-hamiltonian}), one has 
\begin{equation}
Q(H_{F})\coloneqq\hat{H}_{F}=U\hat{H}_{B}U^{-1}\,.
\end{equation}
Since the above relation is a similarity transformation, both operators $%
\hat{H}_{F}$ and $\hat{H}_{B}$ have the same eigenvalues, so from this point
of view $U$ is a mere change of basis in $\mathbb{C}^{2}$. Nevertheless, the
metric $\rho$ is still undefined.

There is a unique realization of the $Q(\zeta)$ algebra, up to the sign of $%
\det \Lambda$, such that the Hamiltonian operators in both quantizations
have the same form, and that realization is 
\begin{equation}
Q\left(\zeta_{k}\right)=\det \Lambda\sqrt{\frac{\hbar}{2}}\sigma_{k}\,.
\end{equation}
In other words, up to a sign, if the $Q$ quantization is realized in the
usual representation by Pauli matrices, $\hat{H}_{F}$ is given by the
operator 
\begin{equation}
\hat{H}_{F}=\frac{\hbar}{2}\boldsymbol{\sigma}\cdot\boldsymbol{F}\,.
\label{HF}
\end{equation}
Thus, starting from this requirement, one fixes the isomorphism $U$ that
will give (\ref{HF}) from (\ref{HB}), so now the inner product $\rho$ is
fixed. As a result, the quantization of $H_{F}$ will provide the operator (%
\ref{HF}). Furthermore, one sees from this procedure that the isomorphism $U$
is unique up to $\rho$-preserving transformations.

So $\hat{H}_{B}$ is the Hamiltonian of a two-level system interacting with a
real magnetic field $\mathbf{B}$, while $\hat{H}_{F}$ is the Hamiltonian of
a two-level system subject to a complex field $\mathbf{F}$. Each Hamiltonian
is hermitian with regard to the underlying inner product, so $\hat{H}_{B}=%
\hat{H}_{B}^{\dagger}$ and $\hat{H}_{F}=\hat{H}_{F}^{+}$. On the
pseudoclassical side, the corresponding symbols are real under their
respective involutions, and the symbols are mapped to each other by the
symplectic transformation $\Lambda$ \eqref{ctc}.

\subsection{The $n=6$ case}

Here we present a new application, which was not considered in \cite%
{baldiotti2021}. Namely, the case of two interacting spins subject to
external magnetic fields. Consider the Lagrange function 
\begin{equation}
L=\frac{\mathrm{i}}{2}\xi _{i}\dot{\xi}_{i}+\frac{\mathrm{i}}{2}\chi _{i}%
\dot{\chi}_{i}-H\left( \xi ,\chi \right) ~,
\end{equation}%
where 
\begin{equation}
H\left( \xi ,\chi \right) =-\frac{\mathrm{i}}{2}\varepsilon _{ijk}\xi
_{i}\xi _{j}B_{k}-\frac{\mathrm{i}}{2}\varepsilon _{ijk}\chi _{i}\chi
_{j}C_{k}+\frac{1}{2}J_{ij}\xi _{i}\chi _{j}  \label{eq:hamiltoniana}
\end{equation}%
and $\left\{ \xi _{i}\right\} _{i=1}^{3}$ and $\left\{ \chi _{i}\right\}
_{i=1}^{3}$ are Grassmann algebra generators which satisfy 
\begin{equation}
\xi _{i}\xi _{j}+\xi _{j}\xi _{i}=0\,,\,\,\chi _{i}\chi _{j}+\chi _{j}\chi
_{i}=0\,,\,\,\,\xi _{i}\chi _{j}=\chi _{j}\xi _{i}\,.
\end{equation}

The conjugate momenta are 
\begin{equation}
\pi _{i}=\frac{\partial L}{\partial \dot{\xi}_{i}}=\frac{\mathrm{i}}{2}\xi
_{i}\,,\,\,\varpi _{i}=\frac{\partial L}{\partial \dot{\chi}_{i}}=\frac{%
\mathrm{i}}{2}\chi _{i}\,,
\end{equation}%
and the canonical Hamilton function becomes 
\begin{align}
H_{c}& =\pi _{i}\dot{\xi}_{i}+\varpi _{i}\dot{\chi}_{i}-L  \notag \\
& =\phi _{i}\dot{\xi}_{i}+\psi _{i}\dot{\chi}_{i}+H\left( \xi ,\chi \right)
~,
\end{align}%
where 
\begin{equation}
\phi _{i}=\pi _{i}-\frac{\mathrm{i}}{2}\xi _{i}\,,\,\,\psi _{i}=\varpi _{i}-%
\frac{\mathrm{i}}{2}\chi _{i}\,,
\end{equation}%
are constraints which satisfy 
\begin{equation}
\left\{ \phi _{i},\phi _{j}\right\} =-\mathrm{i}\delta _{ij}\,,\,\,\left\{
\psi _{i},\psi _{j}\right\} =-\mathrm{i}\delta _{ij}\,,\,\,\left\{ \phi
_{i},\psi _{j}\right\} =0~,
\end{equation}%
according to the Poisson brackets 
\begin{equation}
\left\{ f,g\right\} =\frac{\partial f}{\partial \xi _{i}}\frac{\partial g}{%
\partial \pi _{i}}+\frac{\partial f}{\partial \chi _{i}}\frac{\partial g}{%
\partial \varpi _{i}}-\left( -1\right) ^{P_{f}P_{g}}\frac{\partial g}{%
\partial \xi _{i}}\frac{\partial f}{\partial \pi _{i}}-\left( -1\right)
^{P_{f}P_{g}}\frac{\partial g}{\partial \chi _{i}}\frac{\partial f}{\partial
\varpi _{i}}\,.
\end{equation}%
The consistency conditions for the constraints, $\left\{ \phi
_{i},H_{c}\right\} =\left\{ \psi _{i},H_{c}\right\} =0$, give 
\begin{align}
\dot{\xi}_{i}& =\mathrm{i}\frac{\partial H}{\partial \xi _{i}}=\mathrm{i}%
\left( \mathrm{i}\varepsilon _{ijk}B_{k}+\frac{1}{2}J_{ij}\chi _{j}\right)
=-\varepsilon _{ijk}B_{k}+\frac{\mathrm{i}}{2}J_{ij}\chi _{j}~,  \notag \\
\dot{\chi}_{i}& =\mathrm{i}\frac{\partial H}{\partial \chi _{i}}=\mathrm{i}%
\left( \mathrm{i}\varepsilon _{ijk}C_{k}+\frac{1}{2}J_{ji}\xi _{j}\right)
=-\varepsilon _{ijk}C_{k}+\frac{\mathrm{i}}{2}J_{ji}\xi _{j}~.
\end{align}%
The Dirac brackets are given by 
\begin{equation}
\left\{ f,g\right\} _{D}=\left\{ f,g\right\} -\left\{ f,\phi _{i}\right\}
\left( C^{-1}\right) _{ij}\left\{ \phi _{j},g\right\} -\left\{ f,\psi
_{i}\right\} \left( C^{-1}\right) _{ij}\left\{ \psi _{j},g\right\} ~,
\end{equation}%
where $\left( C^{-1}\right) _{ij}$ is the inverse of $C_{ij}=\left\{ \phi
_{i},\phi _{j}\right\} =\left\{ \psi _{i},\psi _{j}\right\} =-\mathrm{i}%
\delta _{ij}$. The nonvanishing commutators between classical canonical
variables are 
\begin{align}
\left\{ \xi _{i},\xi _{j}\right\} _{D}& =-\mathrm{i}\delta _{ij},\,\,\left\{
\pi _{i},\pi _{j}\right\} _{D}=\frac{\mathrm{i}}{4}\delta _{ij},\,\,\left\{
\xi _{i},\pi _{j}\right\} _{D}=\frac{1}{2}\delta _{ij}~,  \notag \\
\left\{ \chi _{i},\chi _{j}\right\} _{D}& =-\mathrm{i}\delta
_{ij},\,\,\left\{ \varpi _{i},\varpi _{j}\right\} _{D}=\frac{\mathrm{i}}{4}%
\delta _{ij},\,\,\left\{ \chi _{i},\varpi _{j}\right\} _{D}=\frac{1}{2}%
\delta _{ij}~.
\end{align}%
With the help of constraints $\phi $ and $\psi $ we eliminate the momenta
from the classical description, which becomes completely determined by the
generators $\xi $ and $\chi $, and their commutators $\left\{ \xi _{i},\xi
_{j}\right\} _{D}=\left\{ \chi _{i},\chi _{j}\right\} _{D}=-\mathrm{i}\delta
_{ij}$ . Let us consider a linear transformation among the generators, 
\begin{equation}
\xi _{i}^{\prime }=R_{ij}\xi _{j}+R_{ij}^{\prime }\chi _{j}~,\ \chi
_{i}^{\prime }=S_{ij}^{\prime }\xi _{j}+S_{ij}\chi _{j}~.  \label{ct2}
\end{equation}%
For these transformations to be canonical, one has 
\begin{align}
\left\{ \xi _{i}^{\prime },\xi _{j}^{\prime }\right\} _{D}& =-\mathrm{i}%
\delta _{ij}\Rightarrow R_{ik}R_{jk}+R_{ik}^{\prime }R_{jk}^{\prime }=\delta
_{ij}~,  \notag \\
\left\{ \chi _{i}^{\prime },\chi _{j}^{\prime }\right\} _{D}& =-\mathrm{i}%
\delta _{ij}\Rightarrow S_{ik}^{\prime }S_{jk}^{\prime }+S_{ik}S_{jk}=\delta
_{ij}~,  \notag \\
\left\{ \xi _{i}^{\prime },\chi _{j}^{\prime }\right\} _{D}& =0\Rightarrow
R_{ik}S_{jk}^{\prime }+R_{ik}^{\prime }S_{jk}=0~,  \notag \\
\left\{ \chi _{i}^{\prime },\xi _{j}^{\prime }\right\} _{D}& =0\Rightarrow
S_{ik}^{\prime }R_{jk}+S_{ik}R_{jk}^{\prime }=0~.
\end{align}%
The above relations imply that the $6\times 6$ matrix 
\begin{equation}
\Lambda =\left( 
\begin{array}{cc}
R & R^{\prime } \\ 
S^{\prime } & S%
\end{array}%
\right)
\end{equation}%
is orthogonal, 
\begin{equation}
\Lambda \Lambda ^{T}=I_{6}\,,
\end{equation}%
that is, $\Lambda \in O\left( 6,\mathbb{C}\right) $, and $I_{6}$ is the $%
6\times 6$ identity.

Following the general quantization procedure (\ref{quant}), the classical
algebra 
\begin{equation}
\left\{ \xi _{i},\xi _{j}\right\} _{D}=-\mathrm{i}\delta _{ij}\,,\,\,\left\{
\chi _{i},\chi _{j}\right\} _{D}=-\mathrm{i}\delta _{ij}\,,\,\,\left\{ \xi
_{i},\chi _{j}\right\} _{D}=0\,,
\end{equation}%
is mapped to the quantum (anti)commutators using the correspondence 
\begin{align}
\left[ Q\left( \xi _{i}\right) ,Q\left( \xi _{j}\right) \right] _{+}& =%
\mathrm{i}\left\{ \xi _{i},\xi _{j}\right\} _{D}=\delta _{ij}\,,\,\,\left[
Q\left( \chi _{i}\right) ,Q\left( \chi _{j}\right) \right] _{+}=\mathrm{i}%
\left\{ \chi _{i},\chi _{j}\right\} _{D}=\delta _{ij}\,,  \notag \\
& \left[ Q\left( \xi _{i}\right) ,Q\left( \chi _{j}\right) \right] _{-}=%
\mathrm{i}\left\{ \xi _{i},\chi _{j}\right\} _{D}=0\,.
\label{eq:quantum-algebra}
\end{align}%
Note that the commutator grading between operators of classical generators
is the same as the corresponding grading in the classical algebra. A natural
realization of the algebra (\ref{eq:quantum-algebra}) is given by the tensor
product, 
\begin{equation}
Q\left( \xi _{i}\right) =\frac{\sigma _{i}}{\sqrt{2}}\otimes
I_{2}\,,\,\,Q\left( \chi _{i}\right) =I_{2}\otimes \frac{\sigma _{i}}{\sqrt{2%
}}\,,
\end{equation}%
where $I_{2}$ is the $2\times 2$ identity. In matrix form, one has 
\begin{align}
Q\left( \xi _{i}\right) & =\frac{1}{\sqrt{2}}\left( 
\begin{array}{cc}
\sigma _{i} & 0 \\ 
0 & \sigma _{i}%
\end{array}%
\right) \,,Q\left( \chi _{1}\right) =\frac{1}{\sqrt{2}}\left( 
\begin{array}{cc}
0 & I_{2} \\ 
I_{2} & 0%
\end{array}%
\right) \,,  \notag \\
Q\left( \chi _{2}\right) & =\frac{1}{\sqrt{2}}\left( 
\begin{array}{cc}
0 & -\mathrm{i}I_{2} \\ 
\mathrm{i}I_{2} & 0%
\end{array}%
\right) \,,\,\,Q\left( \chi _{3}\right) =\frac{1}{\sqrt{2}}\left( 
\begin{array}{cc}
I_{2} & 0 \\ 
0 & -I_{2}%
\end{array}%
\right) ~.
\end{align}%
Monomials in $\xi $ and $\chi $ are mapped to operators on $\mathcal{L}\left( \mathbb{C%
}^4 \right) $ by total anti-symmetrization of products of $\xi $ and $\chi $%
, and total symmetrization of mixed products $\xi \chi $. Thus, the Hamilton
function (\ref{eq:hamiltoniana}) is mapped to 
\begin{align}
Q\left( H\right) & =-\frac{\mathrm{i}}{2}\varepsilon _{ijk}Q\left( \xi
_{i}\right) Q\left( \xi _{j}\right) B_{k}-\frac{\mathrm{i}}{2}\varepsilon
_{ijk}Q\left( \chi _{i}\right) Q\left( \chi _{j}\right) C_{k}  \notag \\
& +\frac{1}{4}J_{ij}\left( Q\left( \xi _{i}\right) Q\left( \chi _{j}\right)
+Q\left( \chi _{j}\right) Q\left( \xi _{i}\right) \right) ~.
\end{align}%
Therefore, 
\begin{equation}
Q\left( H\right) =\frac{1}{4}\left( B_{k}\sigma _{k}\otimes I_{2}\right) +%
\frac{1}{4}\left( I_{2}\otimes C_{k}\sigma _{k}\right) +\frac{1}{8}%
J_{ij}\left( \sigma _{i}\otimes \sigma _{j}+\sigma _{j}\otimes \sigma
_{i}\right) ~.
\end{equation}%
For the case in which $J_{ij}=J_{i}\delta _{ij}$ (with no summation over $i$
implied), the Hamiltonian operator $Q\left( H\right) $ is 
\begin{equation}
\hat{H}=\frac{1}{4}B_{k}\left( \sigma _{k}\otimes I_{2}\right) +\frac{1}{4}%
C_{k}\left( I_{2}\otimes \sigma _{k}\right) +\frac{1}{4}\sum_{i=1}^{3}J_{i}%
\left( \sigma _{i}\otimes \sigma _{i}\right) \,.
\end{equation}%
Let us split $\hat{H}$ into a free and a interaction part, $\hat{H}=\hat{H}%
_{0}+\hat{H}_{\mathrm{int}}$, so the free part is given by 
\begin{equation}
\hat{H}_{0}=\frac{1}{4}\left( 
\begin{array}{cccc}
B_{3}+C_{3} & B_{1}-\mathrm{i}B_{2} & C_{1}-\mathrm{i}C_{2} & 0 \\ 
B_{1}+\mathrm{i}B_{2} & C_{3}-B_{3} & 0 & C_{1}-\mathrm{i}C_{2} \\ 
C_{1}+\mathrm{i}C_{2} & 0 & -C_{3}+B_{3} & B_{1}-\mathrm{i}B_{2} \\ 
0 & C_{1}+\mathrm{i}C_{2} & B_{1}+\mathrm{i}B_{2} & -C_{3}-B_{3}%
\end{array}%
\right) ~,  \label{h0}
\end{equation}%
while the interaction part is 
\begin{equation}
\hat{H}_{\mathrm{int}}=\frac{1}{4}\left( 
\begin{array}{cccc}
J_{3} & 0 & 0 & J_{1}-J_{2} \\ 
0 & -J_{3} & J_{1}+J_{2} & 0 \\ 
0 & J_{1}+J_{2} & -J_{3} & 0 \\ 
J_{1}-J_{2} & 0 & 0 & J_{3}%
\end{array}%
\right) ~.  \label{hint}
\end{equation}

As is clear from the matrix realization, the operators of the classical
generators $Q\left(\xi_{i}\right)$ and $Q\left(\chi_{i}\right)$ are
hermitian with respect to the canonical inner product in $\mathbb{C}^{4}$.
Thus, whenever classical functions are real with respect to (\ref%
{star-involution}), they are mapped to hermitian operators with respect to
the same product. In this case, the involution maps $\xi_{i}^{\ast}=\xi_{i}$%
, $\chi_{i}^{\ast}=\chi_{i}$, so the Hamiltonian function is real provided
the quantities $B_{i}$, $C_{i}$ and $J_{i}$, $i=1,2,3$, are real.

For the case of a single spin, (the $n=3$ case), the canonical
transformations of the pseudoclassical theory (\ref{ctc}) are given by
transformations $\Lambda\in O\left(3,\mathbb{C}\right)$, and the condition
for the Hamiltonian $\hat{H}=\boldsymbol{\sigma}\cdot\mathbf{F/}$ $2$ to be
pseudo-hermitian is given by \cite{baldiotti2021}, 
\begin{equation}
F^{2}\in\mathbb{R}_{+}\,,\ F^{2}=F_{1}^{2}+F_{2}^{2}+F_{3}^{2}~.
\label{Cond1}
\end{equation}
Condition (\ref{Cond0}) states that the transformations $\Lambda\in O\left(3,%
\mathbb{C}\right)$ preserve the value of $F^{2}$. Thus, the quantum theory
is pseudo-hermitian if and only if there exists a canonical transformation $%
\Lambda $ on the Grassmann symplectic space $M$ which maps a theory with
complex $\mathbf{F}$ to a theory with a real field $\mathbf{B}$ (\ref{FfromB}%
).

One can find a particular solution for the case of two spins which mimics
the properties of the $n=3$ case. The choice $R^{\prime }=S^{\prime }=0$ in
the linear canonical transformation (\ref{ct2})\ preserves the form of the
Hamiltonian (\ref{eq:hamiltoniana}), so that the transformed Hamiltonian is 
\begin{equation}
H^{\prime }\left( \xi ^{\prime },\chi ^{\prime }\right) =-\frac{\mathrm{i}}{2%
}\varepsilon _{ijk}\xi _{i}^{\prime }\xi _{j}^{\prime }F_{k}-\frac{\mathrm{i}%
}{2}\varepsilon _{ijk}\chi _{i}^{\prime }\chi _{j}^{\prime }G_{k}+\frac{1}{2}%
J_{ij}^{\prime }\xi _{i}^{\prime }\chi _{j}^{\prime }
\label{eq:hamiltoniana-complexa}
\end{equation}%
where 
\begin{equation}
F_{k}=(\det R)R_{kl}B_{l}\,,G_{k}=(\det S)S_{kl}C_{l}\,,J_{ij}^{\prime
}=R_{ik}S_{jl}J_{kl}~.
\end{equation}

Quantization will lead to 
\begin{equation}
\hat{H}_{0}=\frac{1}{4}\left[ \left( \boldsymbol{\rho }\mathbf{\cdot G}%
\right) +\left( \boldsymbol{\Sigma }\mathbf{\cdot F}\right) \right] ~,\
\Sigma _{k}=I_{2}\otimes \sigma _{k}\,,\;\rho _{k}=\sigma _{k}\otimes
I_{2}\,,  \label{HL}
\end{equation}%
for the non-interacting part. Its four eigenvalues $E_{\pm }^{\mp }$ are 
\begin{equation}
E_{\pm }^{\mp }=\pm \frac{1}{4}\sqrt{F^{2}+G^{2}\mp 2\sqrt{F^{2}G^{2}}}~.
\label{E}
\end{equation}%
From (\ref{E}) we see that the eigenvalues of the noninteracting two-spin
system are real if 
\begin{equation}
F^{2}\in \mathbb{R}_{+}\text{ and }G^{2}\in \mathbb{R}_{+}~.
\end{equation}%
As expected, in the noninteracting case, in order that the system be
pseudo-hermitian, each piece of the Hamiltonian corresponding to each spin
must be pseudo-hermitian. Moreover, for $J_{ik}=0$, the pseudoclassical
description is simply given by the direct sum $G_{3}\oplus G_{3}$ and the
quantized Hamiltonian is pseudo-hermitian if and only if there are
transformations $R,S\in O(3,\mathbb{C})$ such that 
\begin{equation}
F_{k}=(\det R)R_{kl}B_{l}\,,\ G_{k}=(\det S)S_{kl}C_{l}~,  \label{F2-1}
\end{equation}%
with $\mathbf{B}$ and $\mathbf{C}$ real fields.

The description above highlights a class of solutions that bears many
similarities to the $n=3$ case. However, higher-dimensional cases such as $%
n=6$ admit a much richer description given by the possibility that $%
S^{\prime }$ and $R^{\prime }$ are not zero, for instance. This implies
that, even for the case in which the fields are not related by (\ref{F2-1}),
i.e., in the case of a non-interacting non-hermitian theory, the presence of
a real interaction term ($J_{ik}\in \mathbb{R}$) can lead to a quantized
pseudo-hermitian Hamiltonian, as we will see next.

\section{Physical aspects of two coupled spins}

\label{realization}

As pointed out in the Introduction, one physical advantage of the
generalization for higher dimension is the possibility to model spin
interaction. Here we explore this possibility by analyzing a particular
problem for the $n=6$ case, that is, two coupled spins. We will restrict our
discussion to the case of parallel fields to the spins, when the fields
applied to both spins have the same direction. This case has practical
appeal since it is easier to immerse both spins in the same field and change
the amplitude of the effective field by changing the material where one of
the spins is placed. In what concerns the interaction, a case of special
physical interest is the so-called \textit{Heisenberg interaction} \cite%
{Bal2008}, 
\begin{equation}
\hat{H}_{\mathrm{int}}=\frac{J}{4}\left( \boldsymbol{\Sigma }\mathbf{\cdot }%
\boldsymbol{\rho }\right) \,,\ \left( \boldsymbol{\Sigma }\cdot \boldsymbol{%
\rho }\right) =\sum_{i=1}^{3}\sigma _{i}\otimes \sigma _{i}~,\ J\in \mathbb{R%
}\,.  \label{ih}
\end{equation}%
This interaction has various applications to physical problems, e.g., the
description of the Hubbard model \cite{AshMe76}, the coupling of two quantum
dots \cite{LosDi98} and the modeling of quantum logical gates \cite{Bal2008}%
. Due to the spherical symmetry of the Heisenberg interaction, we can,
without loss of generality, choose the parallel fields in the $z$ direction.
With this choice, and the interaction (\ref{ih}), the total Hamiltonian of
the system becomes%
\begin{equation}
\hat{H}=\hat{H}_{0}+\hat{H}_{\mathrm{int}}=\frac{1}{4}\left[ \rho
_{3}G_{3}+\Sigma _{3}F_{3}+J\left( \boldsymbol{\Sigma }\mathbf{\cdot }%
\boldsymbol{\rho }\right) \right] ~.  \label{Ht0}
\end{equation}%
Observe that for $J=0$, if $G_{3}^{2}$ and $F_{3}^{2}$ are not real, the
above Hamiltonian is non-Hermitian and spin precession is damped \cite%
{baldiotti2021}. In other words, the medium where the spins are placed, as
well as the applied fields, describe a dissipative system. However, as we
will see, by establishing an interaction between the spins it is possible to
bring the system to a pseudo-Hermitian regime and, consequently, suppress
the dissipation. For a real interaction $J>0$, we have the following
eigenvalues for $\hat{H}$ in (\ref{Ht0}): 
\begin{eqnarray}
&&E_{\pm }^{\left( 1\right) }=-J\pm \sqrt{4J^{2}+F_{-}^{2}},\ E_{\pm
}^{\left( 2\right) }=J\pm F_{+}~,  \notag \\
&&F_{\pm }=F_{3}\pm G_{3}~,\ F_{3},G_{3}\in 
\mathbb{C}
~.
\end{eqnarray}%
The system becomes pseudo-hermitian when $\mathrm{Im}(E_{\pm }^{\left(
1\right) })=\mathrm{Im}(E_{\pm }^{\left( 2\right) })=0$, which implies 
\begin{equation}
F_{+},\ F_{-}^{2}\in 
\mathbb{R}
\text{ and }4J^{2}+F_{-}^{2}>0~.  \label{B0}
\end{equation}%
In what follows we consider that these quantities satisfy the above
conditions.

When the fields obey condition (\ref{B0}) it is possible to establish a
isomorphism $U:\mathcal{H}\rightarrow \mathcal{H}_{\rho }$, between the
Hilbert space $\mathcal{H}_{\rho }$, with a metric $\rho ^{-1}=UU^{\dagger }$
and whose the dynamics is governed by (\ref{Ht0}), and the Hilbert space $%
\mathcal{H}=\mathbb{C}^4$ with the canonical metric and whose the dynamics
is governed by a (canonical) Hermitian Hamiltonian%
\begin{equation}
U^{-1}\hat{H}U=\hat{H}_{R}=\frac{1}{4}\left[ \mathbf{B}\cdot \boldsymbol{\
\rho }+\mathbf{C}\cdot \boldsymbol{\Sigma }+\sum_{i=1}^{3}\tilde{J}%
_{i}\left( \Sigma _{i}\rho _{i}\right) \right] =\hat{H}_{R}^{\dag }~,
\label{H-B}
\end{equation}%
with real fields $\mathbf{B}$ and $\mathbf{C}$ and interaction $\mathbf{\ 
\tilde{J}}$. In order to determine $\hat{H}_{R}$ we employ the canonical
limit defined in \cite{baldiotti2021}. Namely, we look for real fields such
that $\det \hat{H}=\det \hat{H}_{R}$ and that when we \textquotedblleft turn
off\textquotedblright\ the complex entries of $\hat{H}$ we obtain the
Hamiltonian $\hat{H}_{R}$. By writing $F_{-}=Re(F_{-})+\mathrm{i}\alpha $,
this canonical limit implies%
\begin{equation}
\det \hat{H}=\det \hat{H}_{R}\text{ and }\lim_{\alpha \rightarrow 0}\hat{H}%
=\lim_{\alpha \rightarrow 0}\hat{H}_{R}~.  \label{cl}
\end{equation}%
It guarantees that, if $\psi ,\phi \in \mathcal{H}$ and $\xi ,\zeta \in 
\mathcal{H}_{\rho }$,%
\begin{equation}
\left\langle \psi ,\phi \right\rangle =\left\langle \xi ,\zeta \right\rangle
_{\rho }~,\ \xi =U\psi ~,\ \zeta =U\phi ~,\ 
\end{equation}%
with%
\begin{equation}
\lim_{\alpha \rightarrow 0}U=I_{6}~,
\end{equation}%
and, consequently, a complete equivalence between the two quantum
descriptions. With this construction, vectors related by $U$ describe the
same physical state.

The following Hamiltonian has the property (\ref{cl}):%
\begin{eqnarray}
\hat{H}_{R} &=&\frac{1}{8}\left[ F_{+}\left( \rho _{3}+\Sigma _{3}\right)
-Re\left( F_{-}\right) \left( \rho _{3}-\Sigma _{3}\right) \right]  \notag \\
&&+\frac{1}{8}\left[ \left( J+E_{+}^{\left( 1\right) }\right)
\sum_{i=1}^{2}\left( \Sigma _{i}\rho _{i}\right) +J\rho _{3}\Sigma _{3}%
\right] ~,  \label{hr}
\end{eqnarray}%
which corresponds to a system with the real fields $\mathbf{B},\mathbf{C}$\
and interaction $\mathbf{\tilde{J}}$, given by%
\begin{eqnarray}
B_{1,2} &=&C_{1,2}=0~,\ B_{3}=\frac{1}{2}\left[ F_{+}+Re\left( F_{-}\right) %
\right] ~,\ C_{3}=\frac{1}{2}\left[ F_{+}-Re\left( F_{-}\right) \right] ~, 
\notag \\
\tilde{J}_{1} &=&\tilde{J}_{2}=\frac{J}{2}+\frac{E_{+}^{\left( 1\right) }}{2}%
=\frac{\sqrt{4J^{2}+F_{-}^{2}}}{2}~,\ \tilde{J}_{3}=J~.  \label{as}
\end{eqnarray}%
We can see that  the system with complex fields is
isomorphic to a system with real fields, but with a non-spherically
symmetric interaction ($\tilde{J}_{3}\neq \tilde{J}_{1,2}$). Note that each
real field involves a combination of both complex fields $\mathbf{F}$ and $%
\mathbf{G}$. Therefore, the present problem is not in the $R^{\prime
}=S^{\prime }=0$ class discussed in the previous section and, even
concerning the non-interaction part, it is essentially quite different from
the $n=3$ case.

Now, knowing $\hat{H}_{R}$, it is straightforward to find the isomorphism $U$%
\ that obeys (\ref{H-B}) and the metric $\rho $,%
\begin{equation}
U=\left( 
\begin{array}{cccc}
1 & 0 & 0 & 0 \\ 
0 & \frac{E_{+}^{\left( 1\right) }+J}{2J} & -\frac{F_{-}}{2J} & 0 \\ 
0 & 0 & 1 & 0 \\ 
0 & 0 & 0 & 1%
\end{array}%
\right) ~,\ \rho =\left( 
\begin{array}{cccc}
1 & 0 & 0 & 0 \\ 
0 & \frac{4J^{2}}{4J^{2}+F_{-}^{2}} & \frac{2JF_{-}}{4J^{2}+F_{-}^{2}} & 0
\\ 
0 & \frac{2JF_{-}}{4J^{2}+F_{-}^{2}} & 1 & 0 \\ 
0 & 0 & 0 & 1%
\end{array}%
\right) ~.  \label{m}
\end{equation}%
The Hamiltonian $\hat{H}$ is Hermitian with respect to $\rho $,%
\begin{equation}
\hat{H}=\rho ^{-1}\hat{H}^{\dagger }\rho =\hat{H}^{+}\,.
\end{equation}

An example involving complex fields is two interacting spins in a
dissipative medium subject to a field in the following form 
\begin{equation}
F_{1,2}=G_{1,2}=0~,\ F_{3}=\frac{1+\mathrm{i}\alpha _{1}}{1+\alpha _{1}^{2}}%
B~,\ G_{3}=\frac{1+\mathrm{i}\alpha _{2}}{1+\alpha _{2}^{2}}B~.  \label{camp}
\end{equation}%
Here $B>0$ is the $z$-component of an external magnetic field. The
quantities $\alpha _{i}\in \mathbb{R}_{+}$, $i=1,2$, represent the so-called
Gilbert damping parameter of the material on the site of spin $i$ \cite%
{baldiotti2021}. For arbitrary values of the parameters in (\ref{camp}), the
Hamiltonian $\hat{H}$\ is in general non-Hermitian, describing a dissipative
system. But, when the parameters satisfy condition (\ref{B0}), the
dissipative process disappears. As a toy model, we can consider the case
where $\alpha _{1}=-\alpha _{2}=\alpha $. In this case, the condition (\ref%
{B0}) is satisfied when%
\begin{equation}
B\leq J\frac{(\alpha ^{2}+1)}{\alpha }~.  \label{B}
\end{equation}%
It shows that the interaction can be used to recover hermiticity of the
Hamiltonian. That is, given a free ($J=0$) non-hermitian Hamiltonian, a real
interaction ($J\in \mathbb{R}$) can make the system pseudo-hermitian.

When the parameters satisfy (\ref{B}) the dynamic of the system, in special
the probability transition, can be computed as%
\begin{equation}
\left\langle \xi ,\exp \left( -\mathrm{i}\hat{H}t\right) \zeta \right\rangle
_{\rho }=\left\langle U^{-1}\xi ,\exp \left( -\mathrm{i}\hat{H}_{R}t\right)
U^{-1}\zeta \right\rangle ~,\ \xi ,\zeta \in \mathcal{H}_{\rho }~,
\end{equation}%
where $U$ and $\hat{H}_{R}$ are given in (\ref{m}) and (\ref{hr}) with%
\begin{eqnarray}
E_{\pm }^{\left( 1\right) } &=&-J\pm 2\sqrt{J^{2}-\frac{B^{2}\alpha ^{2}}{%
(\alpha ^{2}+1)^{2}}}~,~E_{\pm }^{\left( 2\right) }=J\pm \frac{2B}{\alpha
^{2}+1}~,  \notag \\
F_{+} &=&-\mathrm{i}\frac{F_{-}}{\alpha }=\frac{2B}{1+\alpha ^{2}}~.
\end{eqnarray}%
Since $\hat{H}_{R}=\hat{H}_{R}^{\dag }$, \ the evolution is unitary in the
usual sense. Physically it means that when the parameters $B$, $\alpha $ and 
$J$ obey the above condition (\ref{B}), there is no longer any damping in
the system.

Since the Gilbert parameter $\alpha $ describes energy dissipation, it is
assumed strictly positive. That's why we call the above example a toy model.
However, a negative Gilbert damping parameter can be realized, for example,
through a magneto-optical interaction induced by the application of lasers.
Although in this case the spin is not subject to a constant field, the
effective behavior is simply a Landau-Lifshitz-Gilbert equation with a
negative value for the Gilbert parameter (see \cite{Yun} and references
therein). Consequently, the system can be described by a field in the form (%
\ref{camp}) with a negative value of $\alpha _{i}$. Alternatively, a
resonant optical beam can be used to favor a specific spin state, inducing
dissipation by singling out the ground state ($\alpha >0$), or it can be
used to favor energy gain by singling out the excited state ($\alpha <0$).
In the rotating wave approximation, a time-dependent field can be treated as
a constant field in the rotating frame. In \cite{Harter} such a technique is
used to favor the ground state and study the pseudo-Hermitian regime of
two-spin state transitions in the hyperfine levels of $^{6}\mathrm{Li}$
atoms. Therefore, our toy model can become a real system if, by some
technique, the value of the Gilbert parameter was inverted in the second
spin. In this case, besides a possible experimental test of our proposal,
the passage from the non-Hermitian to the pseudo-Hermitian regime can be
used to measure one of the parameters in (\ref{B}). Even though the whole
development has been done considering a constant field $B$, one can consider
adiabatic variations of the magnetic field. Even for an adiabatic variation,
the previous result predicts a drastic change (i.e., non-adiabatic) in the
behavior of the system when $B=J(\alpha ^{2}+1)/\alpha $. Thus, considering $%
B$ to be a known external field, the above result can, in principle, be used
to measure $J$ by knowing $\alpha $, or vice-versa, besides the control of
the dissipative behavior of the system. In addition, the same effects are
expected in a (apparently more realizable) system with real fields $\mathbf{B%
}$, $\mathbf{C}$\ and the axially symmetric coupling $\mathbf{\tilde{J}}$\ (%
\ref{as}).

\section{Final remarks}

\label{final-remarks}

In this work we construct a covariant quantization scheme relating
pseudoclassical mechanics and nonrelativistic Quantum Mechanics with
fermionic degrees of freedom, with the objective of establishing the regimes
within which a pseudo-hermitian Quantum Mechanics is equivalent to ordinary
Quantum Mechanics. This is done in the setting of finite-dimensional
systems, inspired by the results obtained in the context of two-level
systems in \cite{baldiotti2021}.

The quantization map ensures hermiticity of operators, whose symbols are
real functions in the respective pseudoclassical phase space. The
commutativity of the quantization map relates canonical transformations
between symbols to similarity transformations between the corresponding
operators. In particular, the Hamiltonians are real under their classic
involutions, and the corresponding operators are hermitian with respect to
the inner products of the Hilbert spaces whereupon they act.

We have provided a general framework for the quantization of $n$-spin
setups. We then apply this general framework for the case of two interacting
spins, where the interplay between damping and pseudo-hermiticity manifests
itself in the relative intensities among the interaction strength, the
magnetic field intensity and the damping parameter. The suppression of
damping in spin precession under magnetic fields would lead to better spin
manipulation techniques, with manifold technological applications. Coupled
spins, or four level systems, as the one described here, has a wide range of
applications in quantum computation as a physical realization of a two-qubit
gate \cite{Bal2008}. In this sense, the description and control of damping
effects is paramount.

\section*{Acknowledgments:}

R. Fresneda acknowledges the support of São Paulo Research Foundation
(FAPESP), Brazil, with Grant No. 2021/10128-0.

\end{document}